\documentclass{ws-procs975x65}

\def\beq{\begin{equation}}
\def\eeq{\end{equation}}
\def\etal {{\it et al.}}
\newcommand{\unit}[1]{\ {\rm #1}}
\newcommand{\Ref}[1]{Ref.~\refcite{#1}}
\newcommand{\Refs}[1]{Refs.~\refcite{#1}}
\newcommand{\zu}[1]{Fig.~\ref{#1}}
\newcommand{\shiki}[1]{Eq.~(\ref{#1})}
\newcommand{\hyou}[1]{Table~\ref{#1}}
\renewcommand{\Re}{{\rm Re}}
\renewcommand{\Im}{{\rm Im}}
\def\voy{\mathrel{\rlap{\lower0pt\hbox{\hskip1pt{$y$}}}\raise3pt\hbox{$\neg$}}}
\def\ylm#1#2{\voy_{#1}^{#2}}
\newcommand{\Reylm}[2]{\Re[\ylm{#1}{#2}]}
\newcommand{\Imylm}[2]{\Im[\ylm{#1}{#2}]}
\newcommand{\To}{T_{\oplus}}
\newcommand{\omegao}{\omega_{\oplus}}

\newcommand{\rot}{{\rm rot}}
\newcommand{\mpr}{m^{\prime}}
\newcommand{\CCmm}{C^{C}_{m \mpr}}
\newcommand{\CSmm}{C^{S}_{m \mpr}}
\newcommand{\SCmm}{S^{C}_{m \mpr}}
\newcommand{\SSmm}{S^{S}_{m \mpr}}
\def\CC#1{C^{C}_{#1}}
\def\CS#1{C^{S}_{#1}}
\def\SC#1{S^{C}_{#1}}
\def\SS#1{S^{S}_{#1}}

\begin{document}

\title{Higher order test of Lorentz invariance with an optical ring cavity}
\author{Yuta\ Michimura,$^{a*}$ Jake\ Guscott,$^b$ Matthew\ Mewes,$^c$ Nobuyuki\ Matsumoto,$^{def}$\\
Noriaki\ Ohmae,$^g$ Wataru\ Kokuyama,$^h$ Yoichi\ Aso,$^{ij}$ and Masaki\ Ando$^{ai}$}

\address{$^a$Department of Physics, University of Tokyo, Bunkyo, Tokyo 113-0033, Japan\\
$^b$Department of Physics, The University of Adelaide, Adelaide, South Australia 5005, Australia\\
$^c$California Polytechnic State University, San Luis Obispo, California 93407, USA\\
$^d$Frontier Research Institute for Interdisciplinary Sciences, Tohoku University,\\ Sendai, Miyagi 980-8578, Japan \\
$^e$Research Institute of Electrical Communication, Tohoku University,\\ Sendai, Miyagi 980-8577, Japan\\
$^f$PRESTO, Japan Science and Technology Agency, Kawaguchi, Saitama 332-0012, Japan\\
$^g$Department of Applied Physics, University of Tokyo, Bunkyo, Tokyo 113-8656, Japan\\
$^h$National Metrology Institute of Japan, National Institute of Advanced Industrial Science and Technology, Tsukuba, Ibaraki 305-8563, Japan\\
$^i$National Astronomical Observatory of Japan, Mitaka, Tokyo 181-8588, Japan\\
$^j$Department of Astronomical Science, SOKENDAI (The Graduate University for Advanced Studies), Mitaka, Tokyo, 181-8588, Japan\\
$^*$E-mail: michimura@granite.phys.s.u-tokyo.ac.jp}

\begin{abstract}
We have developed an apparatus to search for the higher-order Lorentz violation in photons by measuring the resonant frequency difference between two counterpropagating directions of an asymmetric optical ring cavity. From the year-long data taken between 2012 and 2013, we found no evidence for the light speed anisotropy at the level of $\delta c/c \lesssim 10^{-15}$. Limits on the dipole components of the anisotropy are improved by more than an order of magnitude, and limits on the hexapole components are obtained for the first time. An overview of our apparatus and the data analysis in the framework of the spherical harmonics decomposition of anisotropy are presented. We also present the status of the recent upgrade of the apparatus.
\end{abstract}


\bodymatter


\section{Introduction}
Lorentz invariance underlies all the known physics of fundamental interactions, and is widely accepted as the universal symmetry of nature. All of the great variety of precise experimental tests carried out so far show that Lorentz invariance is valid. However, theoretical attempts to unify the standard model of particle physics and general relativity have led to the possibility that Lorentz invariance may only be approximate. This motivates a further experimental search for Lorentz violation with increasing precision.

Since the experiment done by Michelson and Morley, the search for the anisotropy in the speed of light has been one of the most classical and direct way to test Lorentz invariance in photons. There are two types of searches; even-parity and odd-parity. Even-parity experiments search for the directional dependence in the round-trip speed of light. Michelson-Morley (MM) type experiments are even-parity experiments, and modern versions of MM-type experiments have been carried out by comparing the resonant frequencies of orthogonally aligned microwave cavities~\cite{Nagel} or optical Fabry-P\'{e}rot cavities~\cite{Eisele}. The current best upper limits on even-parity anisotropy obtained from these experiments are at the $10^{-18}$ level~\cite{Nagel}.

On the other hand, odd-parity experiments search for the difference between the speed of light propagating in opposite directions. This search is done by measuring the resonant frequency difference between opposite directions of an asymmetric ring cavity. The best precision before our experiment was at the $10^{-13}$ level~\cite{Baynes}.

To compare the results of the different experiments, the framework of the standard model extension (SME) has been widely used~\cite{SME,HOSME}. Lorentz violating terms in the Lagrangian density of the SME are constructed from a conventional operator with constant tensor coefficients. These unconventional terms are characterized in part by the mass dimension $d$ of the operators. Lorentz violating terms of $d=3$ and $d=4$ are renormalizable, and have been studied extensively both theoretically and experimentally. Nonrenormalizable terms with $d>4$ create a wide variety of higher-order violations. Recently, theoretical works have established phenomenology to search for these higher-order violations experimentally~\cite{HOSME,HOLV}.

Here, we present our odd-parity experiment which reached the $10^{-15}$ sensitivity level, and summarize our limits on higher-order Lorentz violation. We also describe the status of our upgrade of the apparatus

\section{Spherical harmonic decomposition of anisotropy}
A phenomenological way to express the light speed anisotropy is to simply expand the anisotropy with the special harmonics. Using the spherical harmonics
\begin{equation} \label{sphharm}
  Y_{l}^{m}(\theta,\phi) = (-1)^{m} \sqrt{\frac{2 l + 1}{4 \pi} \frac{(l-m)!}{(l+m)!}}  P_l^{m}(\cos{\theta}) e^{i m \phi} ,
\end{equation}
the speed of light can be expanded as
\begin{equation} \label{sphharmexpand}
  c(\theta,\phi) = 1 + \sum_{l=0}^{\infty} \sum_{m=0}^{l} \Re \left[ ( \ylm{l}{m} )^{*} Y_{l}^{m}(\theta,\phi) \right] .
\end{equation}
Here, $\theta \in [0,\pi]$ and $\phi \in [0,2 \pi)$ are the polar angle and the azimuthal angle of the spherical coordinates, respectively. $P_l^{m}$ is the associated Legendre polynomials, and $l$ and $m$ are integers. $\ylm{l}{m}$ are the complex anisotropy spherical coefficients which are zero when Lorentz invariance holds, and $^{*}$ represents the complex conjugate.

The $l=0$ term in \shiki{sphharmexpand} represents the isotropic shift of the speed of light and can be assigned to multiple Lorentz violations, such as dependence of the speed of light on the source velocity, polarization, or wavelength. In this paper, we will neglect these terms because of the two reasons. One reason is because the leading order source velocity modulation only occurs at the period of a year from the revolution of the Earth around the sun. The other reason is because there are strict bounds on the Lorentz violation which cause the birefringence and dispersion.

$l=1,\ 2,\ 3, \cdots$ terms represent the dipole, quadrupole, hexapole, ... components of the anisotropy of the speed of light and can be measured separately by paying attention to the different types of rotational symmetries. $l=2k+1$ terms can be measured with odd-parity experiments, and  $l=2k$ terms can be measured with even-parity, experiments.

Higher-order Lorentz violation is associated with the anisotropy with higher $l$. For example, SME Lorenz violating operators of mass dimension $d=6$ and $d=8$ create the higher-order anisotropy upto $l=2$ and $l=4$, respectively,

This framework of the spherical harmonic decomposition is useful for the analysis independent of the choice of test theory. In this paper, we present our limits on the spherical coefficients $\ylm{l}{m}$ of $l=1,3$ to show the experimental precision in $\delta c/c$. Limits on the higher-order Lorentz violation in the framework of the SME can be found in \Ref{MichimuraPRD}.

\section{Double-pass optical ring cavity}
Symmetric ring cavities in vacuum have sensitivity to the even-parity Lorentz violations, but not to the odd-parity Lorentz violations. This is because the odd-parity terms cancel out over the round-trip path of the cavity (see \shiki{sphharmexpand}). However, a ring cavity will have direct sensitivity to the odd-parity Lorentz violations if the refractive index changes asymmetrically through its path. The resonant frequency of an asymmetric ring cavity shifts when the Lorentz invariance is violated. Since the signs of the resonant frequency shift are opposite between the clockwise and counterclockwise directions of the ring cavity, measuring the resonant frequency difference between opposite directions gives the Lorentz violation signal.

\begin{figure}[t]
  \begin{center}
    \includegraphics[width=\hsize]{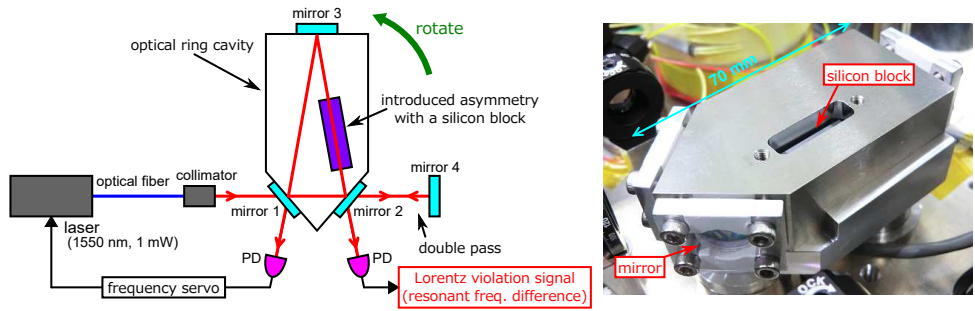}
    \caption{Left: Schematic of our setup. Right: Asymmetric optical ring cavity.} \label{schematic}
  \end{center}
\end{figure}

The schematic of our setup is shown in \zu{schematic}. Our ring cavity is constructed from three mirrors and the silicon block is inserted in one side of the triangle to introduce asymmetry. The large refractive index of silicon (measured value $n=3.69$ at wavelength $\lambda = 1550$ nm) gives better sensitivity to the odd-parity Lorentz violations.

The resonant frequency comparison is done with a double pass configuration. The 1550 nm laser beam is first injected into the ring cavity in counterclockwise direction. Using the error signal from the reflection from mirror1, the laser frequency is locked to the counterclockwise resonant frequency. The cavity transmitted beam from mirror2 is then reflected back into the cavity in the clockwise direction. The second error signal from the reflection from mirror2 is proportional to the resonant frequency difference, and in this signal we search for the Lorentz violation.

The whole optics and the laser source is put on a turntable and rotated to modulate the Lorentz violation signal. Positive and reverse rotations of $420^{\circ}$ were repeated alternately to avoid twisting of the power supply cables and the signal extraction cables. The rotation speed was $\omega_{\rot}=30^{\circ} \unit{/sec}$ (0.083 Hz). For more details of the apparatus, see \Refs{MichimuraPRL,Thesis}.

\section{Limits on higher-order Lorentz violation}

The ring cavity was rotated approximately 1.7 million times during the data acquisition done at the University of Tokyo for 393 days between August 2012 and September 2013. To test for Lorentz violation, we first consider a decomposition of the measured resonant frequency difference $\delta \nu/\nu$ into harmonics of $\omega_\rot$,
\begin{equation}
  \frac{\delta \nu}{\nu} = 
  \sum_{m > 0} \big[ C_m \cos{(m \omega_\rot t)} + S_m \sin{(m \omega_\rot t)} \big] \ .
  \label{demod1}
\end{equation}
The amplitudes $C_m$ and $S_m$ vary at harmonics of Earth's sidereal frequency $\omegao$ and can be expanded as
\begin{align}
  C_m &= \sum_{\mpr \ge 0} \big[ \CCmm \cos{(\mpr\alpha)} + \CSmm \sin{(\mpr\alpha)} \big]\ ,
  \nonumber \\ 
  S_m &= \sum_{\mpr \ge 0} \big[ \SCmm \cos{(\mpr\alpha)} + \SSmm \sin{(\mpr\alpha)} \big]\ ,
  \label{demod2}
\end{align}
where $\alpha = \omegao \To$ is the right ascension of the local zenith\ \cite{HOSME}. In our analysis, we search for the dipole and the hexapole components of the anisotropy. Thus, we restrict attention to $m=1, 3$ and $\mpr = 0, 1, 2, 3$.

\begin{table}[t]
\begin{center}
\begin{tabular}{c}
\begin{minipage}{0.6\hsize}
\renewcommand{\arraystretch}{1.1}
\tbl{Constraints on the modulation amplitudes. All values are in units of $10^{-15}$. }
{\begin{tabular}{@{}cc||cc@{}}
    \toprule
    Amplitude & Measurement & Amplitude & Measurement \\
    \colrule
    $\CC{10}$ & $-0.1\pm1.0$ & $\CC{30}$ & $0.12\pm0.46$ \\ 
    $\SC{10}$ & $ 0.2\pm1.0$ & $\SC{30}$ & $0.15\pm0.46$ \\
    $\CC{11}$ & $-0.6\pm1.4$ & $\CC{31}$ & $-0.79\pm0.64$ \\
    $\CS{11}$ & $-1.2\pm1.4$ & $\CS{31}$ & $-1.1\pm0.65$ \\ 
    $\SC{11}$ & $-0.3\pm1.4$ & $\SC{31}$ & $-0.48\pm0.64$ \\ 
    $\SS{11}$ & $ 1.0\pm1.4$ & $\SS{31}$ & $-0.51\pm0.65$ \\ 
    $\CC{12}$ & $-0.9\pm1.4$ & $\CC{32}$ & $-1.1\pm0.65$ \\ 
    $\CS{12}$ & $-0.2\pm1.4$ & $\CS{32}$ & $ 0.57\pm0.65$ \\ 
    $\SC{12}$ & $-0.1\pm1.4$ & $\SC{32}$ & $-0.46\pm0.65$ \\ 
    $\SS{12}$ & $ 1.0\pm1.4$ & $\SS{32}$ & $ 0.21\pm0.65$ \\ 
    $\CC{13}$ & $-0.8\pm1.4$ & $\CC{33}$ & $ 0.40\pm0.65$ \\ 
    $\CS{13}$ & $ 0.2\pm1.4$ & $\CS{33}$ & $ 0.16\pm0.65$ \\ 
    $\SC{13}$ & $-0.5\pm1.4$ & $\SC{33}$ & $-0.36\pm0.64$ \\ 
    $\SS{13}$ & $ 0.6\pm1.4$ & $\SS{33}$ & $ 0.75\pm0.65$ \\
    \botrule
\end{tabular}}
\label{modresult}
\end{minipage}
\begin{minipage}{0.4\hsize}
\renewcommand{\arraystretch}{1.1}
\tbl{Spherical coefficients with $1\sigma$ uncertainties determined from this work. All values are in units of $10^{-15}$.} 
{\begin{tabular}{@{}ccc@{}}
\toprule
Coefficient & Measurement \\
\hline
$  \ylm{1}{0}$ & $0.4 \pm 4.4$ \\
$\Reylm{1}{1}$ & $-5.7 \pm 6.3$ \\
$\Imylm{1}{1}$ & $-3.2 \pm 6.2$ \\
$  \ylm{3}{0}$ & $0.1 \pm 1.9$ \\
$\Reylm{3}{1}$ & $2.9 \pm 2.2$ \\
$\Imylm{3}{1}$ & $-3.2 \pm 2.1$ \\
$\Reylm{3}{2}$ & $2.1 \pm 1.8$ \\
$\Imylm{3}{2}$ & $1.5 \pm 1.8$ \\
$\Reylm{3}{3}$ & $-0.2 \pm 2.2$ \\
$\Imylm{3}{3}$ & $-0.7 \pm 2.2$ \\
\botrule
\end{tabular}}
\label{ylmresult}
\end{minipage}
\end{tabular}
\end{center}
\end{table}

We first demodulated the data at $m \omega_\rot$ to extract the amplitudes $C_m$ and $S_m$ in \shiki{demod1} for each rotation. Day-long time series data for $C_m$ and $S_m$ are then fitted to \shiki{demod2} with the least-squares method to extract the modulation amplitudes $\CCmm$, $\CSmm$, $\SCmm$, and $\SSmm$ for each day. The weighted averages of each modulation amplitude over the whole 393 day period are listed in \hyou{modresult}.

The calculation done in \Ref{Thesis} gives the relationship between these modulation amplitudes and the spherical coefficients. Each of the modulation amplitudes is a linear combination of the spherical coefficients which are dependent upon the laser wavelength, the orientation, the length and refractive index of each arm of the cavity, and the colatitude ($\chi = 54.3^{\circ}$) of the laboratory.

The results are summarized in \hyou{ylmresult}. All the measured coefficients are consistent with zero at $2 \sigma$, and we concluded that there's no significance for Lorentz violation. Our limits on $l=1$ dipole coefficients are at the $6\times10^{-15}$ level, which was more than an order of magnitude better than the previous best experiment~\cite{Baynes}. Our limits on $l=3$ hexapole coefficients are at the $2\times10^{-15}$ level, and are the first limits to our knowledge.

\section{Upgrade status}
\zu{sensitivity} shows the fractional frequency noise spectra of our setup. The blue curve shows the noise when the cavity is stationary, and the red curve shows the noise when the cavity is rotating. The hundredfold increase in the noise is suspected to come from the vibration of the turntable. Also, the frequency resolution of the spectrum when rotating is restricted to the rotational frequency (0.083 Hz) because of the alternative rotation.

To overcome these limitations, we are currently upgrading the apparatus, as summarized in \zu{upgrade}. As for the optics, we reduced the height of the optical axis with respect to the breadboard from 2 in. to 0.5 in. to reduce the vibration sensitivity. Optical mounts for the new apparatus are now fixed directly on the breadboard to make the optics semi-monolithic. The data logger is now also fixed on the turntable to rotate together with the optics, and the data is transferred wirelessly to acquire data continuously. The AC power is now supplied with a slip ring to make the continuous rotation possible. The continuous rotation will help to increase the stability and also the frequency resolution.

At the time of writing, we have tested the new wireless data logger with the power supply from the slip ring. We have also assembled the semi-monolithic optics and made the first measurements of the stationary noise spectra. Detailed study on the noise is now ongoing.

\begin{figure}[t]
  \begin{center}
    \begin{tabular}{c}
\begin{minipage}{0.5\hsize}
  \begin{center}
    \includegraphics[width=\hsize]{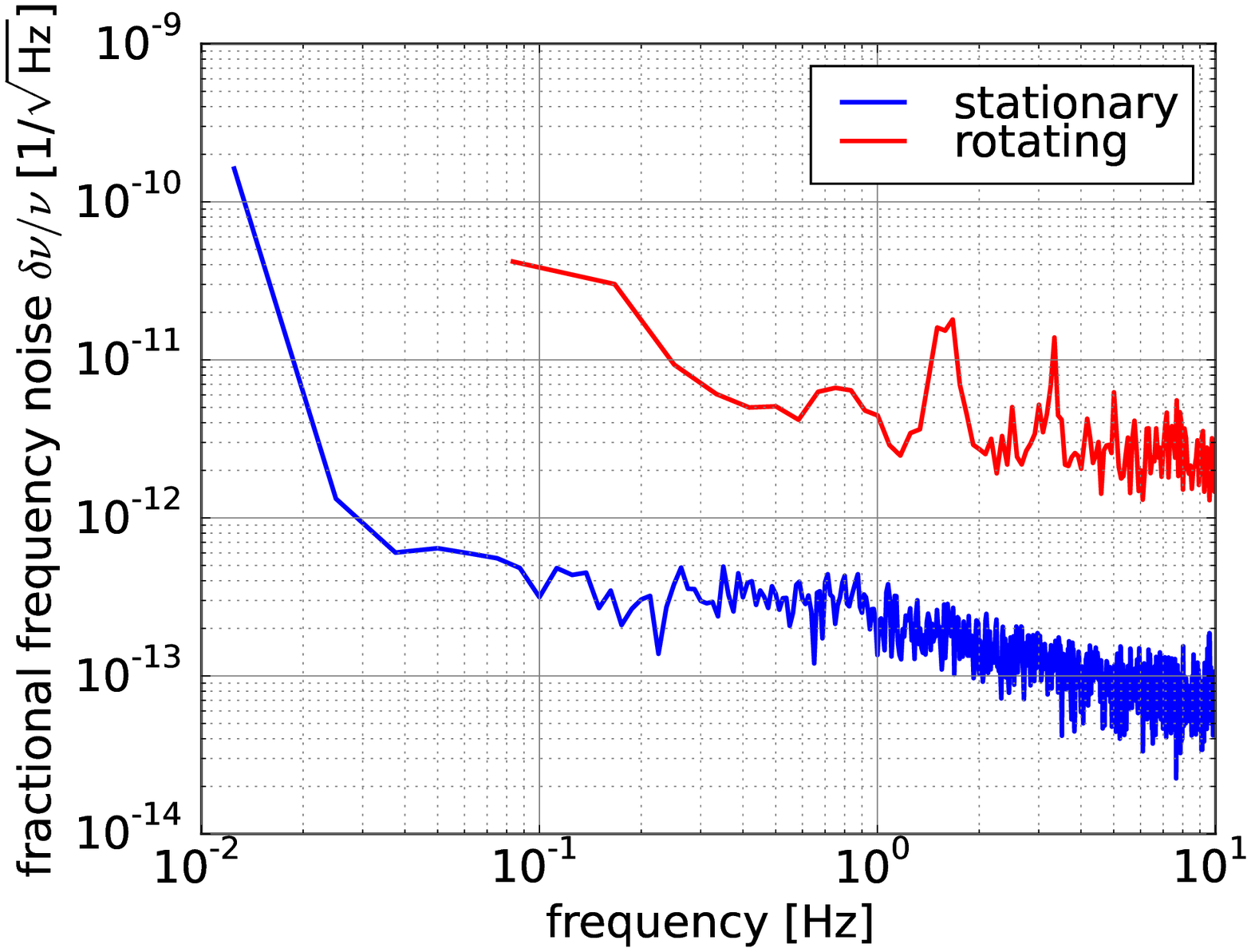}
    \caption{Noise spectra of the previous setup.} \label{sensitivity}
  \end{center}
\end{minipage}
\begin{minipage}{0.5\hsize}
  \begin{center}
    \includegraphics[width=0.9\hsize]{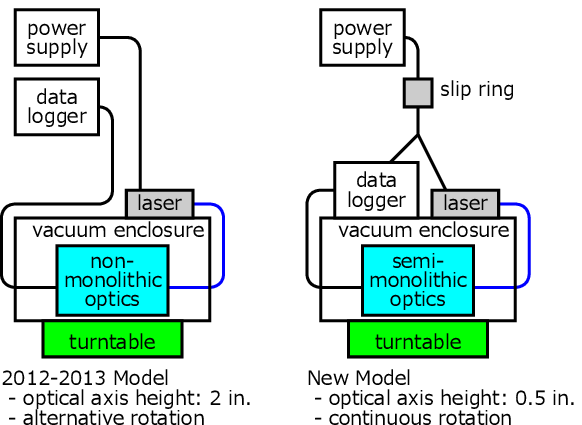}
    \caption{Summary of the upgrade.} \label{upgrade}
  \end{center}
\end{minipage}
    \end{tabular}
  \end{center}
\end{figure}

\section{Conclusion}
We searched for the higher-order Lorentz violation in photons using a double-pass optical ring cavity. No significant evidence for the anisotropy in the speed of light was found at the level of $10^{-15}$. Our limits on the dipole components are more than an order of magnitude better than the previous best limits, and our limits on the hexapole components are the first constraints.

The upgrade of the apparatus is underway, and the upgraded apparatus has the potential to improve the sensitivity by 2 orders of magnitude.

\section*{Acknowledgments}
We thank Shigemi Otsuka for manufacturing some of the mechanical parts of the apparatus. Jake Guscott acknowledges financial support from The Commonwealth Government of Australia through Endeavour Scholarships and Fellowships. This work was supported by JSPS Grant-in-Aid for Young Scientists (A) No. 15H05445.

\end{document}